\documentclass[12pt]{article}
\usepackage[xdvi]{graphicx}
\renewcommand{\theequation}{\thesection.\arabic{equation}} 
\makeatletter
\@addtoreset{equation}{section}
\makeatother

\newcommand{\bea}{\begin{eqnarray}}
\newcommand{\eea}{\end{eqnarray}}

\begin{document}
\begin{titlepage}
\begin{flushright} 
HIP-2007-19/TH \\
arXiv:0704.2490
\end{flushright}

\vspace*{2cm}

\begin{center}{\large\bf 
Supersymmetry breaking by constant superpotentials
and O'Raifeartaigh model
in warped space
}
\end{center}
\vspace{1.5cm}

\begin{center}
{\bf Nobuhiro Uekusa}
\footnote{E-mail: nobuhiro.uekusa@helsinki.fi}
\end{center}

\vspace{0.2cm}

\begin{center}
{\it High Energy Physics Division, 
Department of Physical Sciences, University of Helsinki  
and Helsinki Institute of Physics, 
P.O. Box 64, FIN-00014 Helsinki, Finland}
\end{center}

\vspace{1cm}

\begin{abstract}
Supersymmetry breaking together 
by constant boundary superpotentials 
and by the O'Raifeartaigh model
is studied 
in a warped space model.
It is shown that the contribution of
constant boundary superpotentials enables
the moduli of chiral supermultiplets
to be stabilized
and that 
the vacuum at the stationary point
has zero cosmological constant
in a wide region of parameters.

\end{abstract}
\end{titlepage}
\newpage
\renewcommand{\theequation}{\thesection.\arabic{equation}} 
\section{Introduction}

Supersymmetric models with extra dimensions
has attracted interest.
Since extra dimensions and supersymmetry have not been discovered,
these must be invisible
at low scales. 
One of the simple ways to compactify extra dimensions and 
break supersymmetry simultaneously
is the Scherk-Schwarz mechanism. 
It is known that supersymmetry breaking by the Scherk-Schwarz mechanism 
\cite{Scherk:1978ta}\cite{
Scherk:1979zr}
is equivalent 
to the supersymmetry 
breaking by 
constant superpotentials 
in flat bulk space \cite{Marti:2001iw}--%
\cite{Biggio:2002rb}. 
These two scenarios generate the same mass spectrum. 

The equivalence between supersymmetry breakings by 
the Scherk-Schwarz mechanism and constant
superpotentials has been discussed also in warped space
\cite{Altendorfer:2000rr}--%
\cite{Correia:2006pj},
particularly in the Randall-Sundrum background \cite{Randall:1999ee}.
Here the more fundamental question has been examined,
i.e., whether
supersymmetry can be 
broken by the Scherk-Schwarz mechanism in Randall-Sundrum
background.
From the viewpoint of supergravity, 
the answer seems to be negative.
This statement means that 
the Scherk-Schwarz mechanism cannot be only the source of
supersymmetry breaking in Randall-Sundrum background.
On the other hand, the degrees of freedom of the Scherk-Schwarz twist
may exist in Randall-Sundrum background 
if other supersymmetry-breaking sources are taken into
account \cite{Abe:2004ar}.
Thus it would be important to clarify the effects of
Scherk-Schwarz twists or constant superpotentials
in systems with additional sources.

In the previous papers \cite{Maru:2006id}\cite{Maru:2006ji},
we have shown that
a warped space model with a constant boundary superpotential
is an efficient model both to break supersymmetry
and to stabilize the radius
when a hypermultiplet, a compensator and a radion multiplet are 
taken into account.
We presented possible additional supersymmetry-breaking sources of
$F$-term and $D$-term 
to cancel  the cosmological constant.
The resulting
soft scalar mass, gravitino mass and radion mass as well as 
zero 
cosmological constant
all gave evidence that this model is phenomenologically viable.
In this model, 
the sectors of constant superpotentials and
additional supersymmetry breaking 
are decoupled.
It would be worth to work with
systems where these sectors are coupled
because if constant superpotentials are allowed only
in the case with supersymmetry breaking in additional sectors, 
they may be mixing each other.

In this Letter
we study supersymmetry breaking in a warped space model with
constant boundary superpotentials, a hypermultiplet, 
a compensator, a radion multiplet and boundary chiral supermultiplets.
Equations of motion are solved together for these fields.
We take into account
the mass parameter $c$ for the hypermultiplet and
a superpotential of 
O'Raifeartaigh model \cite{O'Raifeartaigh:1975pr}
for the boundary chiral supermultiplets.
If the hypermultiplet is decoupled, the model reduces to
ordinary O'Raifeartaigh model.
There is a flat direction of the chiral supermultiplets.
In the presence of the hypermultiplet,
it is shown that the flat direction 
is lifted due to the mixing
of the equations of motion.
Then we show that 
a modulus of the hypermultiplet
remains unfixed for zero constant superpotentials
and that it is stabilized  
in the case with nonzero constant superpotentials for large negative
$c$.
In other words, the additional stabilization 
of moduli can be developed when the sector of
constant superpotentials is coupled to a system 
with spontaneous supersymmetry breaking.

\section{Model}\label{sc:model}
We consider a five-dimensional supersymmetric model of 
a single hypermultiplet 
and three chiral supermultiplets
on the Randall-Sundrum background \cite{Randall:1999ee}
whose metric is 
\bea
ds^2 = e^{-2R\sigma}\eta_{\mu\nu}dx^\mu dx^\nu +R^2 dy^2, 
\quad ~~
\sigma(y)\equiv k|y|, 
\eea
where $R$ is the radius of $S^1$ of the orbifold $S^1/Z_2$, 
$k$ is the 
curvature of the five-dimensional Anti-de-Sitter (AdS$_5$) space, 
and the angle of $S^1$ 
is denoted by $y\,(0 \le y \le \pi)$. 
In terms of superfields for 
four-dimensional $N=1$ supersymmetry,
our Lagrangian 
 is \cite{Marti:2001iw}\cite{Maru:2006id}
\begin{eqnarray}
\!\!\!
{\cal L} &\!\!\!=\!\!\!& \int 
 d^2 \theta d^2\bar{\theta}~
\frac{1}{2}\, \varphi^\dag \varphi (T+T^\dag) 
e^{-(T+T^\dag)\sigma}
(\Phi^\dag \Phi + \Phi^c \Phi^{c\dag} - 6M_5^3) 
\nonumber \\
&& + \int d^2 \theta 
\left[
\varphi^3 e^{-3T \sigma} \left\{
\Phi^c \left[
\partial_y - \left( \frac{3}{2} - c \right)T \sigma' 
\right] \Phi + W_c
\right\} + \textrm{H.c.}
\right] 
\nonumber
\\
 &&
+\delta(y) \left[
   \int 
 d^2\theta d^2\bar{\theta} ~
 \varphi^\dagger \varphi
  (\Phi_1^\dagger \Phi_1 +\Phi_2^\dagger \Phi_2 + \Phi_3^\dagger \Phi_3)
  +\left\lbrace
  \int d^2\theta ~
 \varphi^3 W(\Phi_1,\Phi_2,\Phi_3)+ \textrm{H.c.}\right\rbrace
 \right] ,
\nonumber
\\
\label{lag}
\end{eqnarray}
where the compensator chiral supermultiplet $\varphi$, 
and the radion chiral supermultiplet $T$ are denoted as
\begin{eqnarray}
\varphi = 1 + \theta^2 F_{\varphi}, ~~~~~~
T=R + \theta^2 F_T ,
\end{eqnarray}
respectively and the chiral supermultiplets representing 
the hypermultiplet are denoted as  
\begin{eqnarray}
 \Phi =\phi +\theta^2 F , ~~~~~~ \Phi^c =\phi^c +\theta^2 F^c .
\end{eqnarray}
The $Z_2$ parity is assigned to be even for 
$\Phi$ and odd for $\Phi^c$.
The derivative with respect to $y$ is denoted by $'$, 
such as $\sigma'\equiv d\sigma/dy$. 
The five-dimensional Planck mass is denoted as $M_5$. 
We assume the constant (field independent) 
superpotential localized at the fixed points $y=0$ and $y= \pi$, 
\begin{eqnarray}
W_c \equiv 2M_5^3 (w_0 \delta(y) + w_\pi \delta(y-\pi)), 
\label{eq:boundary_pot}
\end{eqnarray}
where $w_{0}$ and $w_{\pi}$ are dimensionless constants.
In the Lagrangian (\ref{lag}), the last line is
the O'Raifeartaigh model
coupled to the compensator.
The three chiral supermultiplets are denoted as
\begin{eqnarray}
 \Phi_i = \phi_i +\theta^2 F_i , ~~~~i=1,2,3 .
\end{eqnarray}
which are confined at $y=0$.
The superpotential $W$ is given by 
\begin{eqnarray}
  W(\Phi_i)=\lambda(\Phi_1^2-\mu^2)\Phi_2+m\Phi_1\Phi_3
\end{eqnarray}
where $\lambda,\mu,m$ are real parameters.

As the part of the Lagrangian (\ref{lag}) containing auxiliary 
components is relevant to extra dimensions,
we extract the part 
\begin{eqnarray}
\!\!\!
{\cal L}_{\textrm{\scriptsize aux}} &\!\!\!=\!\!\!& 
\left[
\frac{1}{2}e^{-2R\sigma}(2RF^\dag F + F_T F^\dag \phi 
+ F_T^\dag F \phi^\dag)
\right. \nonumber \\
&& \left. + \left\{
\frac{1}{2}e^{-2R\sigma}(2R \phi^\dag F 
+ F_T(\phi^\dag \phi -3M_5^3))
(F_{\varphi}^\dag - F_T^\dag \sigma)
+ {\rm h.c.} \right\} 
+(\phi \leftrightarrow \phi^c) \right] \nonumber \\
&&+e^{-2R\sigma}R(\phi^\dag \phi + \phi^c \phi^{c\dag} - 6M_5^3)
(F_{\varphi}^\dag - F_T^\dag \sigma)
(F_{\varphi} - F_T \sigma) \nonumber \\
&& + \left[
3e^{-3R\sigma}(F_{\varphi} - F_T \sigma)
\left\{
\phi^c \left[ \partial_y 
-\left( \frac{3}{2} - c \right) R \sigma' \right]
\phi + W_c 
\right\} \right. \nonumber \\
&&\left. +e^{-3R\sigma}\left\{
F^c \left[ \partial_y 
-\left( \frac{3}{2} - c \right) R \sigma' \right]\phi
+ \phi^c \left[ \partial_y 
-\left( \frac{3}{2} - c \right) R \sigma' \right]F 
\right. \right. \nonumber \\
&& \left. \left.  
-\phi^c \left( \frac{3}{2} - c \right) F_T \sigma' \phi 
\right\} +{\rm H.c.}
\right]
\nonumber
\\
&&
+
 \delta (y)
  \left[|F_\varphi|^2|\phi_i|^2
  +|F_i|^2
 +
 \left\lbrace
  F_\varphi \phi_iF_i^\dagger
   +3F_\varphi W
   +W_i F_i
  +\textrm{H.c.}
  \right\rbrace\right]
\label{aux}
\end{eqnarray}
where the summation over $i$ is taken.
The derivatives of the superpotential are denoted
as $W_i\equiv \partial W/\partial\phi_i$, $i=1,2,3$.
The Lagrangian (\ref{aux}) gives the following equations of 
motion for auxiliary fields: 
\begin{eqnarray}
F &\!\!\!=&\!\!\! -\frac{e^{-R\sigma}}{R}
\left[
-\partial_y \phi^{c\dag} 
+ \left( \frac{3}{2} + c \right)R \sigma' \phi^{c\dag}
+\frac{\phi}{2M_5^3}W_c^\dagger 
+\frac{\phi}{6M_5^3}(3W^\dagger-\phi_i^\dagger W_i^\dagger)\delta(y) 
\right. \nonumber \\
&\!\!\!&\!\!\! \left. +\frac{1}{6M_5^3}\phi^\dag \phi 
\partial_y \phi^{c\dag} 
+\frac{1}{3M_5^3}\phi^{c\dag} \phi 
\partial_y \phi^{\dag} 
-\frac{1}{6M_5^3}\phi^\dag \phi 
\phi^{c\dag}\left( \frac{9}{2} -c \right)
R \sigma'
\right], 
\label{Feom} \\
F^c &\!\!\!=&\!\!\! -\frac{e^{-R\sigma}}{R}
\left[
\partial_y \phi^{\dag} 
- \left( \frac{3}{2} - c \right)R \sigma' \phi^{\dag}
+\frac{\phi^c}{2M_5^3}W_c^\dagger
+\frac{\phi^c}{6M_5^3}(3W^\dagger-\phi_i^\dagger W_i^\dagger)\delta(y) \right. \nonumber \\
&\!\!\!&\!\!\! \left. +\frac{1}{6M_5^3}\phi^c \phi^\dag 
\partial_y \phi^{c\dag} 
+\frac{1}{3M_5^3}\phi^{c\dag} \phi^c 
\partial_y \phi^{\dag} 
-\frac{1}{6M_5^3}\phi^c \phi^\dag 
\phi^{c\dag} \left( \frac{9}{2} -c \right)
R \sigma' 
\right],
 \label{Fceom} \\
F_{\varphi} &\!\!\!=&\!\!\! -\frac{e^{-R\sigma}}{R}
\left[
-\frac{1}{6M_5^3}\phi^\dag \partial_y \phi^{c\dag} 
-\frac{1}{3M_5^3}\phi^{c\dag}\partial_y \phi^\dag 
+\frac{1}{6M_5^3} \phi^\dag \phi^{c\dag}
\left( \frac{9}{2} -c \right)R\sigma'
-\frac{1}{2M_5^3}W_c^\dagger 
\right. \nonumber \\
&\!\!\!&\!\!\! \left.
-\frac{3(1-2R\sigma)}{r}\phi^{c\dag} 
\partial_y \phi^\dag -\frac{3(1-2R\sigma)}{r}W_c^\dagger 
+\frac{1-2R\sigma}{r}
\phi^{c\dag}\phi^\dag \left(\frac{3}{2} -c \right)R\sigma'
\right.
\nonumber
\\
 &\!\!\!&\!\!\!\left.
 +\left(-{1\over 6M_5^3}
   -{(1-2R\sigma)^2\over r}\right)
   (3W^\dagger-\phi_i^\dagger W_i^\dagger)
  \delta(y)
\right], \label{Fpeom} \\
F_T &\!\!\!=&\!\!\! -\frac{e^{-R\sigma}}{r}
\left[
6\phi^{c\dag} \partial_y \phi^\dag -2\phi^{c\dag}\phi^\dag 
\left(\frac{3}{2} -c \right)R\sigma' +6W_c^\dagger
\right.
\nonumber
\\
 &\!\!\!&\!\!\! \left.
 +2(1-2R\sigma)(3W^\dagger-\phi_i^\dagger W_i^\dagger)\delta(y)
\right], \label{FTeom} \\
F_i &\!\!\!=&\!\!\! \frac{e^{-R\sigma}}{R}
\phi_i\left[
-\frac{1}{6M_5^3}\phi^\dag \partial_y \phi^{c\dag} 
-\frac{1}{3M_5^3}\phi^{c\dag}\partial_y \phi^\dag 
+\frac{1}{6M_5^3} \phi^\dag \phi^{c\dag}
\left( \frac{9}{2} -c \right)R\sigma'
-\frac{1}{2M_5^3}W_c^\dagger 
\right. \nonumber \\
&\!\!\!&\!\!\! \left.
-\frac{3(1-2R\sigma)}{r}\phi^{c\dag} 
\partial_y \phi^\dag -\frac{3(1-2R\sigma)}{r}W_c^\dagger 
+\frac{1-2R\sigma}{r}
\phi^{c\dag}\phi^\dag \left(\frac{3}{2} -c \right)R\sigma'
\right.
\nonumber
\\
 &\!\!\!&\!\!\!\left.
 +\left(-{1\over 6M_5^3}
   -{(1-2R\sigma)^2\over r}\right)
   (3W^\dagger-\phi_i^\dagger W_i^\dagger)
  \delta(y)
\right]
   -W_i^\dagger
 \label{f1}
\end{eqnarray}
where we have defined
\begin{eqnarray*}
 r \equiv \phi^\dag \phi + \phi^{c\dag} \phi^c - 6M_5^3 .
\end{eqnarray*}
In Eq.(\ref{Feom})
a partial integration has been performed.
The Lagrangian (\ref{aux}) are written as
\begin{eqnarray}
 \!\!\!\!\!\!\!
{\cal L}_{\textrm{\scriptsize aux}}
&\!\!\!=\!\!\!& e^{-3R\sigma}\left[
 (-\partial_y \phi^c+\left({3\over 2}+ c\right)R\sigma'\phi^c)F
 \right. 
+(\partial_y\phi-\left(\frac{3}{2}-c\right)R\sigma'\phi)F^c 
\nonumber
\\
&&\!\!\!\!\! 
+\left(3\phi^c(\partial_y\phi-\left(\frac{3}{2} -c\right) 
R\sigma'\phi) +3W_c +(3W-\phi_i W_i) \delta(y)\right) F_\varphi
\nonumber
\\
&&\!\!\!\!\! \left.
-\left(3\sigma\phi^c(\partial_y\phi-\left(\frac{3}{2} -c\right) 
R\sigma'\phi)+3\sigma W_c
 +\phi^c\left({3\over 2}-c\right)\sigma'\phi\right)F_T
\right] 
 -|W_i|^2 \delta(y).
\label{auxap}
\end{eqnarray}
with $F,F^c,F_\varphi,F_T$
given in Eqs.(\ref{Feom})--(\ref{FTeom}).

\section{Moduli and potential 
}\label{sc:oo}

\subsection{The O'Raifeartaigh sector (no hypermultiplet)}

We begin with examining moduli
in the part of the O'Raifeartaigh model coupled to the compensator. 
If the hypermultiplet is absent,
the Lagrangian (\ref{auxap}) becomes 
\begin{eqnarray} 
 {\cal L}_{\textrm{\scriptsize aux}}
 &\!\!\!=\!\!\!&{e^{-R\sigma}\over 6M_5^3}\, 4\sigma(1-R\sigma)
   \left|3W-\phi_i W_i\right|^2(\delta(y))^2
 -\left|W_i\right|^2\delta(y)
 =
 -\left|W_i\right|^2\delta(y)
 \label{nos}
\end{eqnarray}
where we used $\sigma(\delta(y))^2=0$.
The Lagrangian (\ref{nos}) is the same as in 
the O'Raifeartaigh model without the compensator. 
The solution of $\phi_1$ is
\begin{eqnarray}
 \phi_1=\left\{\begin{array}{l}
  0 ~\textrm{or}~ \pm\sqrt{\mu^2-{m^2/ (2\lambda^2)}}
 \qquad \textrm{for}~ \mu^2 > {m^2/ (2\lambda^2)}
\\
 0 \qquad\qquad\qquad\qquad\qquad\qquad
   \textrm{for}~ \mu^2 < {m^2/ (2\lambda^2)}
\\
 \end{array}\right.
 \equiv \underline{\phi_1} .
 \label{xo}
\end{eqnarray}
The other fields $\phi_2,\phi_3$ only need to satisfy a single equation
\begin{eqnarray}
 2\lambda \phi_1\phi_2+ m\phi_3=0 ,
 \label{phi23}
\end{eqnarray}
and one (or two) of $\phi_2$ and $\phi_3$ 
is undetermined.

From Eqs.(\ref{xo}), (\ref{phi23}) and (\ref{nos}), 
the potential is obtained as
\begin{eqnarray}
 V =-\int_0^\pi dy \,{\cal L}_{\textrm{\scriptsize aux}}
   =|\lambda(\underline{\phi_1}^2-\mu^2)|^2 +|m\underline{\phi_1}|^2
   \ge 0
 \label{po}
\end{eqnarray}
where $\underline{\phi_1}$ is given in (\ref{xo}).
In order to be consistent with the Randall-Sundrum background,
the parameters $\lambda$ and $m$ must be zero. 

In this pure boundary chiral supermultiplet
case, 
the compensator has no 
effects on the
potential and the background solution.

\subsection{Mixing of the two sectors ($W_c=0$)
}\label{sc:mix}

We next examine the background, potential and moduli 
in the case with nonzero hypermultiplet and chiral supermultiplets.
In the model without chiral supermultiplets,
we found that the hypermultiplet solution for $W_c=0$  
is \cite{Maru:2006id} 
\begin{eqnarray}
&&\phi=N_2\exp\left[\left({3\over 2}-c\right)R\sigma\right]
   \equiv 
   \underline{\phi}, 
 \label{phibar}
\\
&&\phi^c=0 
\end{eqnarray}
where $N_2$ is an overall complex constant for the 
flat direction $\phi$.
From this situation, it is one possibility that
a simplest nontrivial solution may exist for $W_c=\phi^c=0$ 
even with chiral supermultiplets. 
In this section, we consider the case $W_c=\phi^c=0$.

From the Lagrangian (\ref{auxap}), 
the equations of motion ($W_c=\phi^c=0$) are 
\begin{eqnarray*}
  &&\!\!\!\!\!
  (\partial_y\phi-\left({3\over 2}-c\right)R\sigma'\phi)
   \left[-\left({3\over 2}-c\right)R\sigma'\right]
-e^{4R\sigma}\partial_y\{(\partial_y\phi
  -\left({3\over 2}-c\right)R\sigma'\phi)
   e^{-4R\sigma}\}
\nonumber
\\
 && \qquad\qquad
  +|3W-\phi_i W_i|^2 (\delta(y))^2 
    {(1-2R\sigma)^2\over r^2}\phi=0 
  \qquad\qquad\qquad \textrm{for}~ \phi^\dagger ,
 \label{phieom}
\\
 &&\!\!\!\!\!
  (3W-\phi_i W_i)\delta(y)
 \left[
  -{1\over 3M_5^3}\partial_y\phi^\dagger
 +{1\over 6M_5^3}\phi^\dagger\left({9\over 2}-c\right)R\sigma'
  -{3\over r}\partial_y\phi^\dagger
  +{1\over r}\phi^\dagger\left({3\over 2}-c\right)R\sigma'\right]
=0 
\nonumber
\\ &&
 \qquad\qquad\qquad\qquad\qquad\qquad\qquad
 \qquad\qquad\qquad\qquad\qquad\qquad\quad
  \textrm{for}~ \phi^c{}^\dagger ,
\\
 &&\!\!\!\!\!
  (3W-\phi_i W_i)(\delta(y))^2 \left(-{1\over R}\right)
   (-{1\over 6M_5^3}-{1\over r})(m\phi_3)^\dagger
\nonumber
\\
 && \qquad\qquad
  -((2\lambda \phi_1\phi_2+m\phi_3)(2\lambda \phi_2)^\dagger
  +\lambda(\phi_1^2-\mu^2)(2\lambda \phi_1)^\dagger 
  +m\phi_1 m^\dagger)\delta(y)=0 
\nonumber
\\ &&
 \qquad\qquad\qquad\qquad\qquad\qquad\qquad
 \qquad\qquad\qquad\qquad\qquad\qquad\quad
 \textrm{for}~ \phi_1^\dagger,
\\
 && \!\!\!\!\!
 (3W-\phi_i W_i)(\delta(y))^2\left(-{1\over R}\right)
   (-{1\over 6M_5^3}-{1\over r})(-2\lambda\mu^2)^\dagger 
  -(2\lambda \phi_1\phi_2+ m\phi_3)(2\lambda \phi_1)^\dagger \delta(y)
  =0 
\nonumber
\\
 &&
  \qquad\qquad\qquad\qquad\qquad\qquad\qquad
 \qquad\qquad\qquad\qquad\qquad\qquad\quad
 \textrm{for} ~ \phi_2^\dagger ,
 \\
 &&\!\!\!\!\!
 (3W-\phi_i W_i)(\delta(y))^2\left(-{1\over R}\right)
  (-{1\over 6M_5^3}-{1\over r})(m\phi_1)^\dagger
 -(2\lambda \phi_1\phi_2+m\phi_3)m^\dagger \delta(y)=0 
\nonumber
\\
&&
 \qquad\qquad\qquad\qquad\qquad\qquad\qquad
 \qquad\qquad\qquad\qquad\qquad\qquad\quad
  \textrm{for} ~\phi_3^\dagger .
\end{eqnarray*}
These equations of motion reduce to
\begin{eqnarray*}
 &&2\lambda \phi_1\phi_2+m\phi_3=0, 
~~~~~\qquad\qquad\qquad
 -2\lambda\mu^2\phi_2+m\phi_1\phi_3=0, \\
 &&\lambda(\phi_1^2-\mu^2)(2\lambda \phi_1)^\dagger 
  +m\phi_1 m^\dagger=0, 
~~~~~~
 \partial_y\phi-\left({3\over 2}-c\right)R\sigma'\phi=0 ,
\end{eqnarray*} 
where $3W-\phi_i W_i=-2\lambda\mu^2\phi_2+m\phi_1\phi_3$.
The first equation above is the same as Eq.(\ref{phi23}). 
The second equation gives an additional constraint to 
$\phi_i$.
As a result, 
there are four equations to determine
the four variables $\phi_i$ and $\phi$.
We find the solution (for $W_c=0$) 
\begin{eqnarray}
 \phi=   \underline{\phi},~~
 \phi^c=0 ,~~
 \phi_1=\underline{\phi_1},~~
 \phi_2=0,~~
 \phi_3=0 . 
  \label{solwo}
\end{eqnarray}
Thus the fields $\phi_i$ are all determined
unlike no hypermultiplet case
in the previous section.
Still $\underline{\phi}$ includes unfixed $N_2$ and $R$.

From the solution (\ref{solwo}) and the Lagrangian (\ref{lag}),
the potential itself is seen to be the same as 
in the O'Raifeartaigh model 
which is given in Eq.(\ref{po}).
In this potential, the moduli $N_2$ and the radius $R$
are not stabilized.
In other words, even if the two sectors have been mixed,
there still exist moduli for the $W_c=0$ case.

\subsection{Moduli stabilization with $W_c\neq 0$ 
}\label{sc:w}
Now we study the case with a
nonzero constant superpotential $W_c$. 
We assume $|w_0|\sim |w_\pi|\equiv w\ll 1$
and work out perturbative solutions of the equations of motion 
for $\phi$, $\phi^c$ and $\phi_i$ similarly to analysis in \cite{Maru:2006id}.
To allow possible discontinuities of the $Z_2$-odd field 
$\phi^c$ across the fixed points $y=0$ and $y= \pi$, we define 
\begin{equation}
\phi^c(x,y) \equiv \hat \epsilon(y) \chi^c(x,y), 
\qquad 
\hat \epsilon(y)\equiv 
\left\{
\begin{array}{cc}
+1, & 0<y<\pi \\
-1, & -\pi<y<0
\end{array}
\right.
, 
\label{eq:odd_field}
\end{equation}
where $\chi^c(x,y)$ is a parity even function with possibly 
nonvanishing value at $y=0, \pi$.  
Up to ${\cal O}(w)$,
the solution for $\phi$ is found to be $\phi=\underline{\phi}$
which is given in Eq.(\ref{phibar}).
Using this solution $\phi=\underline{\phi}$
and
examining $(\delta(y))^2$ terms 
in the equation of motion
for $\phi^\dagger$ derived from the Lagrangian (\ref{auxap}), 
we find that 
\begin{eqnarray*}
 3W-\phi_i W_i \propto w_0 ,
\end{eqnarray*}
which is of order of ${\cal O}(w)$.
As for singular terms,
we use the following 
identity valid as a result of a properly regularized 
calculation: 
\begin{equation}
\delta(y) (\hat \epsilon(y))^2 = \frac{1}{3} \delta(y), 
\qquad 
\delta(y-\pi) (\hat \epsilon(y))^2 = \frac{1}{3} \delta(y-\pi). 
\label{eq:delta_epsilon2}
\end{equation}
This 
respects
the relation $2\delta(y)=d\epsilon(y)/dy$.

From the Lagrangian (\ref{auxap}),
the equation of motion for $\phi^\dagger$ is identical to 
Eq.(\ref{phieom})
up to the first order of $w$.
The equation of motion for $\phi^c{}^\dagger$ up to ${\cal O}(w)$ is
\begin{eqnarray}
 &&
 (-\partial_y\phi^c+\left({3\over 2}+c\right)R\sigma'\phi^c)
 \left[\left({3\over 2}+c\right)R\sigma'
  +{1\over 3M_5^3}\phi\partial_y\phi^\dagger
  -{1\over 6M_5^3}\phi^\dagger\phi\left({9\over
				   2}-c\right)R\sigma'\right]
\nonumber
\\
&&
 -e^{4R\sigma}\partial_y\{(-\partial_y\phi^c+\left({3\over 2}+c\right)
 R\sigma'\phi^c)e^{-4R\sigma}\left[
 -1+{1\over 6M_5^3}\phi^\dagger\phi\right]\}
\nonumber
\\
&&
+(\partial_y\phi -\left({3\over 2}-c\right)R\sigma'\phi)
 \left[{1\over 3M_5^3}\phi^c\partial_y\phi^\dagger
   -{1\over 6M_5^3}\phi^c\phi^\dagger
   \left({9\over 2}-c\right)R\sigma'\right]
\nonumber
\\
&&-e^{4R\sigma}\partial_y\{
   (\partial_y\phi-\left({3\over 2}-c\right)R\sigma'\phi)e^{-4R\sigma}
  {1\over 6M_5^3}\phi^c\phi^\dagger\}
\nonumber
\\
&& 
 +(3\phi^c(\partial_y\phi-\left({3\over 2}-c\right)R\sigma'\phi))
 \left[
  -{1\over 3M_5^3}\partial_y\phi^\dagger
   +{1\over 6M_5^3}\phi^\dagger\left({9\over 2}-c\right)R\sigma'
 \right.
\nonumber
\\
&&  \left.
   -{3(1-2R\sigma)\over r}\partial_y\phi^\dagger
  +{1-2R\sigma\over r}
   \phi^\dagger \left({3\over 2}-c\right)R\sigma'\right]
\nonumber
\\
 &&-e^{4R\sigma}\partial_y\{
  (3\phi^c(\partial_y\phi-\left({3\over 2}-c\right)R\sigma'\phi)
  +3W_c+(3W-\phi_i W_i)\delta(y))e^{-4R\sigma}
  \left[-{1\over 6M_5^3}\phi^\dagger\right]\}
\nonumber
\\
 &&-(3\sigma\phi^c(\partial_y\phi-\left({3\over 2}-c\right)
  R\sigma'\phi)+\phi^c\left({3\over 2}-c\right)R\sigma'\phi)
  {1\over r}\left[
   6\partial_y\phi^\dagger
 -2\phi^\dagger\left({3\over 2}-c\right)R\sigma'\right]
\nonumber
\\
 &&
=0
\end{eqnarray}
where $\hat{\epsilon}\delta(y)=0$ is used
and $r=\phi^\dagger\phi-6M_5^3+{\cal O}(w^2)$ should be taken. 
The equation of motion for $\phi_1^\dagger$ 
up to ${\cal O}(w)$ is
\begin{eqnarray}
 &&(-\partial_y\phi^c+\left({3\over 2}+c\right)R\sigma'\phi^c)
   {\phi \over 6M_5^3}(m\phi_3)^\dagger \delta(y)
 +(\partial_y\phi-\left({3\over 2}-c\right)R\sigma'\phi)
   {\phi^c\over 6M_5^3}(m\phi_3)^\dagger \delta(y)
\nonumber
\\
 &&
+(3\phi^c(\partial_y\phi-\left({3\over 2}-c\right)R\sigma'\phi)
  +3W_c+(3W-\phi_i W_i)\delta(y))
  (-{1\over 6M_5^3}-{1\over r})(m\phi_3)^\dagger\delta(y)
\nonumber
\\
 &&
 -\phi\phi^c\left({3\over 2}-c\right)R\sigma'\,{1\over r}\,(m\phi_3)^\dagger
  \,2\delta(y)
\nonumber
\\
 &&-((2\lambda \phi_1\phi_2+m\phi_3)(2\lambda \phi_2)^\dagger
   +\lambda(\phi_1^2-\mu^2)(2\lambda \phi_1)^\dagger
  +m\phi_1m^\dagger)\delta(y) =0 .
\end{eqnarray}
The equation of motion for $\phi_2^\dagger$ up to ${\cal O}(w)$ is
\begin{eqnarray}
 &&(-\partial_y\phi^c+\left({3\over 2}+c\right)R\sigma'\phi^c)
   {\phi \over 6M_5^3}(-2\lambda\mu^2)^\dagger \delta(y)
\nonumber
\\
 &&
 +(\partial_y\phi-\left({3\over 2}-c\right)R\sigma'\phi)
   {\phi^c\over 6M_5^3}(-2\lambda\mu^2)^\dagger \delta(y)
\nonumber
\\
 &&
+(3\phi^c(\partial_y\phi-\left({3\over 2}-c\right)R\sigma'\phi)
  +3W_c+(3W-\phi_i W_i)\delta(y))
  (-{1\over 6M_5^3}-{1\over r})(-2\lambda\mu^2)^\dagger\delta(y)
\nonumber
\\
 &&
 -\phi\phi^c\left({3\over 2}-c\right)
  R\sigma'\,{1\over r}\,(-2\lambda\mu^2)^\dagger
  \,2\delta(y)
 -(2\lambda \phi_1\phi_2+m\phi_3)(2\lambda \phi_1)^\dagger\delta(y) =0 .
\end{eqnarray}
The equation of motion for $\phi_3^\dagger$ up to ${\cal O}(w)$ is
\begin{eqnarray}
 &&(-\partial_y\phi^c+\left({3\over 2}+c\right)R\sigma'\phi^c)
   {\phi \over 6M_5^3}(m\phi_1)^\dagger \delta(y)
 +(\partial_y\phi-\left({3\over 2}-c\right)R\sigma'\phi)
   {\phi^c\over 6M_5^3}(m\phi_1)^\dagger \delta(y)
\nonumber
\\
 &&
+(3\phi^c(\partial_y\phi-\left({3\over 2}-c\right)R\sigma'\phi)
  +3W_c+(3W-\phi_i W_i)\delta(y))
  (-{1\over 6M_5^3}-{1\over r})(m\phi_1)^\dagger\delta(y)
\nonumber
\\
 &&
 -\phi\phi^c\left({3\over 2}-c\right)
  R\sigma'\,{1\over r}\,(m\phi_1)^\dagger
  \,2\delta(y)
 -(2\lambda \phi_1\phi_2+m\phi_3)m^\dagger\delta(y) =0 .
\end{eqnarray}
From the equations of motion for $\phi^\dagger$ and $\phi^c{}^\dagger$,
it is seen that the hypermultiplet has
the same bulk solutions as in the model without 
the boundary chiral supermultiplets.
The solutions are given for generic values of the bulk 
mass parameter $c$ ($\not=3/2, 1/2$) as \cite{Maru:2006id}
\begin{eqnarray} 
 \phi&\!\!\!=\!\!\!&\underline{\phi}  ,
 \label{solchi}
\\
 \phi^c&\!\!\!=\!\!\!& 
\hat{\epsilon}\,
   {(X+1)^{(5/2-c)/(3-2c)}\over X}
   \left[c_1 +c_2 (X+1)^{-(1-2c)/(3-2c)}\left(X+{3-2c\over
					 1-2c}\right)\right]
 \label{solchic}
\end{eqnarray}
where $c_1$ and $c_2$ are constants of integration. 
We have changed a variable from $y$ to a dimensionless 
variable 
$X\equiv \underline{\phi}^\dagger\underline{\phi}/(6M_5^3)-1$.
The remaining parts of the equations of motion give boundary conditions.
The $\partial_y\delta(y)$, $\partial_y\delta(y-\pi)$ terms 
of the equation of motion for $\phi^c{}^\dagger$ gives rise to
the boundary conditions 
\begin{eqnarray}
 \chi^c\bigg|_{y=0}
  &\!\!\!=\!\!\!&-\left({1\over 2X}\,
 \underline{\phi}^\dagger\left(w_0+{3W-\phi_i W_i\over
     6M_5^3}\right)\right)\bigg|_{y=0} ,
 \label{bc1}
\\
 \chi^c\bigg|_{y=\pi}
  &\!\!\!=\!\!\!&\left({1\over
		  2X}\,\underline{\phi}^\dagger
		  w_\pi\right)\bigg|_{y=\pi} .
 \label{bc2}
\end{eqnarray}
From the equations of motion for $\phi_2^\dagger$ and $\phi_3^\dagger$,
the boundary conditions are 
\begin{eqnarray}
 &&2\lambda \phi_1\phi_2+m\phi_3=0 ,
 \label{xym}
\\
 &&(-\partial_y\phi^c+\left({3\over 2}+c\right)R\sigma'\phi^c)
   \underline{\phi}\delta(y)
  +(6M_5^3 w_0+3W-\phi_i W_i)(\delta(y))^2
   \left(-{\underline{\phi}^\dagger\underline{\phi}\over r}\right)
\nonumber
\\
 &&\quad
  -\underline{\phi}\hat{\epsilon}\chi^c
   \left({3\over 2}-c\right)R\sigma'{6M_5^3\over r}
  2\delta(y)=0 .
 \label{yz}
\end{eqnarray}
In Eq.(\ref{yz}), the $(\delta(y))^2$ terms give the same boundary 
condition as Eq.(\ref{bc1}).
The other terms lead to
\begin{eqnarray}
 \delta(y)\hat{\epsilon}
 \left[-\partial_y\chi^c+\left({3\over 2}+c\right)R\sigma'\chi^c
   -\chi^c\left({3\over 2}-c\right)R\sigma'
  {12M_5^3\over r} \right] =0 .
 \label{bc3}
\end{eqnarray}
This equation gives the boundary condition for $\chi^c$ at $y=0$.
Eq.(\ref{bc1}) gives the boundary condition for $(3W-\phi_i W_i)$ rather than 
for $\chi^c$ at $y=0$.
The boundary condition for $\chi^c$ at $y=\pi$ is given 
by Eq.(\ref{bc2}).
Finally, the equation of motion for $\phi_1^\dagger$ becomes
\begin{eqnarray}
 \phi_1=\underline{\phi_1}
 \label{bc4}
\end{eqnarray}
subject to the boundary conditions (\ref{xym}) and (\ref{yz}).

The boundary conditions given above are solved in the following.
Firstly we calculate the constants of integration $c_1$ and $c_2$.
Substituting the bulk solution (\ref{solchic})
into the boundary conditions (\ref{bc2}) and (\ref{bc3}) one obtains
\begin{eqnarray*}
 &&c_1+c_2 \hat{N}^{(2c-1)/(3-2c)}{1-2c\over 3-2c}=0 ,
\\
 &&c_1+c_2(\hat{N}e^{(3-2c)Rk\pi})^{(2c-1)/(3-2c)}
   (\hat{N}e^{(3-2c)Rk\pi}-{2\over 3-2c})
  ={N_2^\dagger e^{(3/2-c)Rk\pi}w_\pi
 \over 2(\hat{N}e^{(3-2c)Rk\pi})^{(5/2-c)/(3-2c)}} ,
\end{eqnarray*}
where we defined a dimensionless parameter 
$\hat{N}\equiv |N_2|^2/(6M_5^3)$. 
These equations are solved as
\begin{eqnarray}
  c_1&\!\!\!=\!\!\!&
  {(2c-1)N_2^\dagger \hat{N}^{-(5/2-c)/(3-2c)}e^{-Rk\pi}
  \over e^{(2c-1)Rk\pi}((3-2c)\hat{N}e^{(3-2c)Rk\pi}-2)
   +2c-1}{w_\pi\over 2} ,
 \label{c1}
\\
  c_2&\!\!\!=\!\!\!&
  {(3-2c)N_2^\dagger \hat{N}^{-(3/2+c)/(3-2c)}e^{-Rk\pi}
  \over e^{(2c-1)Rk\pi}((3-2c)\hat{N}e^{(3-2c)Rk\pi}-2)
   +2c-1}{w_\pi\over 2} .
 \label{c2} 
\end{eqnarray}
The coefficients $c_1$ and $c_2$ are independent of $w_0$.
Lastly, from Eqs.(\ref{bc1}), (\ref{xym}) and (\ref{bc4}),
$\phi_i$ are solved as
\begin{eqnarray}
  \phi_1&\!\!\!=\!\!\!&\underline{\phi_1} ,
\\
 \phi_2&\!\!\!=\!\!\!&-{6M_5^3\over 2\lambda (\mu^2+\underline{\phi_1}^2)}
    \left({(3-2c)(1-\hat{N})e^{-Rk\pi}w_\pi
     \over e^{(2c-1)Rk\pi}((3-2c)\hat{N}e^{(3-2c)Rk\pi}-2)
    +2c-1}
   -w_0\right) ,
\\
 \phi_3&\!\!\!=\!\!\!& {6M_5^3 \underline{\phi_1}\over m (\mu^2+\underline{\phi_1}^2)}
    \left({(3-2c)(1-\hat{N})e^{-Rk\pi}w_\pi
     \over e^{(2c-1)Rk\pi}((3-2c)\hat{N}e^{(3-2c)Rk\pi}-2)
    +2c-1}
   -w_0\right) ,
 \label{zsol}
\end{eqnarray}
where $\underline{\phi_1}$ is given in Eq.(\ref{xo}).
Obviously, the solutions
(\ref{solchi}),(\ref{solchic})
and (\ref{c1})--(\ref{zsol})
 include the result in the previous section where $w_0=w_\pi=0$.
As in the previous section,
$\phi_i$ are determined unambiguously.

Here we would like to stress that
dependence of the above solutions on $w_0,w_\pi$ are
different from that of the case 
decoupled to 
boundary chiral supermultiplets.
When boundary chiral supermultiplets are absent,
the boundary conditions for $\chi^c$ 
are given in Eq.(\ref{bc2}) and
Eq.(\ref{bc1}) with zero ($3W-\phi_i W_i$).
At $y=0$ the boundary condition  includes $w_0$.
Then $c_1$ and $c_2$ depend on $w_0$ and $w_\pi$.
Even for zero $w_\pi$, there exists a nontrivial solution for $\phi^c$. 
On the other hand, when the boundary chiral supermultiplets are coupled,
Eq.(\ref{bc1}) is interpreted as a boundary condition 
for $\phi_i$ or more concretely for ($3W-\phi_i W_i$).
The three fields $\phi_i$  are
solved for the three equations 
(\ref{bc1}), (\ref{xym}) and (\ref{bc4}).
The boundary conditions for $\chi^c$ are Eqs.(\ref{bc2})
and (\ref{bc3}). They do not
include $w_0$.
Thus the coefficients $c_1$ and $c_2$ are independent of $w_0$.
In obtaining a nontrivial solution for $\phi^c$,
it is required that $w_\pi$ is at least nonzero. 

We have 
solved the equations of motion.
We can now calculate the potential.
By inserting the solutions into the Lagrangian (\ref{aux}) and 
integrating over the extra dimension $y$, 
we obtain the potential as a function of the radius $R$ 
and the complex normalization parameter $N_2$ 
\begin{eqnarray}
 V&\!\!\!=\!\!\!&{k\over 2M_5^3}\int_0^{\pi}
   dy \bigg\{
  -2c_2^\dagger\hat{N}^{5/2-2c+2/(3-2c)}
     e^{((3-2c)(5/2-2c)+2)R\sigma}
\nonumber
\\
  && +\left({3\over 2}+c+(3-2c)
  \left(-{5\over 2}+2c-\left[3(\hat{N}
 e^{(3-2c)R\sigma}-1)\right]^{-1}\right)\right)
   \chi^c{}^\dagger \bigg\}
    {\underline{\phi}^\dagger \widetilde{W}\over 2}e^{-4R\sigma} 
\nonumber
\\
 && +|\lambda(\underline{\phi_1}^2-\mu^2)|^2+|m\underline{\phi_1}|^2
\end{eqnarray}
where $\widetilde{W}\equiv W_c+(W-\phi_i W_i/3)\delta(y)$ 
and  $c_2$ is given in Eq.(\ref{c2}) 
for generic values of $c$.
This form of the potential is similar to the 
decoupled model
\cite{Maru:2006id}.
Performing integration and 
using the boundary conditions (\ref{bc2}) and 
(\ref{bc3}) lead to the potential 
\begin{eqnarray}
 V&=& -N_2^\dagger k
  c_2^\dagger\hat{N}^{5/2-2c+2/(3-2c)}
     \left(\widetilde{w}_0+w_\pi e^{((3-2c)^2-2)Rk\pi}\right)
\nonumber\\
  &&+{|N_2|^2 k\widetilde{w}_0^2\over 4(1-\hat{N})}
       \left(-4c^2+12c-6+{3-2c\over 3(1-\hat{N})}\right)
\nonumber\\
 &&+{|N_2|^2 k w_\pi^2 e^{-(1+2c)Rk\pi}
     \over 4(\hat{N}e^{(3-2c)Rk\pi}-1)}
     \left(-4c^2+12c-6-{3-2c\over 3(\hat{N}e^{(3-2c)Rk\pi}-1)}\right) 
\nonumber
\\
 && +|\lambda(\underline{\phi_1}^2-\mu^2)|^2+|m\underline{\phi_1}|^2 
\end{eqnarray}
where we defined $\widetilde{w}_0\equiv w_0+(3W-\phi_i W_i)/(6M_5^3)$.
Using Eqs.(\ref{c1})--(\ref{zsol}) for
$c_1,c_2,\phi_i$, we find the potential
\begin{eqnarray}
 V&\!\!\!=\!\!\!&{(6M_5^3)kw_\pi^2\over 4}
  \bigg\{
  -{2(3-2c)\hat{N}^{7/2-2c+(1/2-c)/(3-2c)}
    e^{((3-2c)^2-3)Rk\pi}
   \over e^{(2c-1)Rk\pi}((3-2c)\hat{N}e^{(3-2c)Rk\pi}-2)+2c-1}
\nonumber
\\
 &&
 +\hat{N}(1-\hat{N})
  \left({(3-2c)e^{-Rk\pi}\over 
     e^{(2c-1)Rk\pi}((3-2c)\hat{N}e^{(3-2c)Rk\pi}-2)+2c-1}\right)^2
\nonumber
\\
 &&\quad \times
  \left(-4c^2+12c-6+{3-2c\over 3(1-\hat{N})}
   -2\hat{N}^{5/2-2c+(1/2-c)/(3-2c)}\right)
\nonumber
\\
 &&
  +{e^{-(1+2c)Rk\pi}\over \hat{N}e^{(3-2c)Rk\pi}-1}
 \left(-4c^2+12c-6-{3-2c\over 3(\hat{N}e^{(3-2c)Rk\pi}-1)}\right)
  \bigg\}
\nonumber
\\
 && +|\lambda(\underline{\phi_1}^2-\mu^2)|^2+|m\underline{\phi_1}|^2 .
 \label{nowo}
\end{eqnarray}
This potential is independent of $w_0$ 
as it is seen from the fact that $\widetilde{w}_0$ is proportional to 
$w_\pi$ subject to Eq.(\ref{bc1}).

We move on
the stabilization of the radius $R$
and the modulus $N_2$.
For simplicity, we consider the case where $-c\gg 1$ and the 
constant $N_2$ is real.
Then the potential becomes
\begin{eqnarray}
 \!\!\!\!\!
 V\approx 
  -(6M_5^3)kw_\pi^2 c^2 ~
   {(-\hat{N}^2-\hat{N}^{-2c}/(2c)^2)e^{-2Rk\pi}+\hat{N}-1
   \over (\hat{N}-1)(\hat{N}-e^{2cRk\pi})}
  +|\lambda(\underline{\phi_1}^2-\mu^2)|^2+|m\underline{\phi_1}|^2 .
 \label{vw}
\end{eqnarray}
We need to require the stationary condition for both modes 
$R$ and $N_2$
\begin{eqnarray}
 {\partial V\over \partial R}=0 
\textrm{~~and~~} 
 {\partial V\over \partial \hat{N}}=0 .
\end{eqnarray}
The former condition $\partial V/\partial R=0$ leads to 
\begin{eqnarray}
 e^{-2Rk\pi}\approx {(\hat{N}-1)^2\over \hat{N}^2}
 \label{vv1}
\end{eqnarray}
whereas the latter condition gives 
\begin{eqnarray}
 0\approx \left(\hat{N}^3+{\hat{N}^{-2c}\over 2c^2}\right)
        {(\hat{N}-1)^2\over \hat{N}^2}
   +c(-\hat{N}^2+\hat{N}-1)e^{2cRk\pi} .
  \label{vv2}
\end{eqnarray}
From Eqs.(\ref{vv1}) and (\ref{vv2}), 
we find that the stationary condition is satisfied for infinite
radius
and that the modulus $N_2$ is stabilized at
\begin{eqnarray}
 N_2\approx \sqrt{6M_5^3} .
\end{eqnarray}
It is important to notice that 
this stabilization originates from the terms proportional to $w_\pi^2$
in the potential (\ref{vw}). 
Only when
a constant superpotential at $y=\pi$ is nonzero,
the modulus of the hypermultiplet is stabilized.
At the stationary point, the potential is
\begin{eqnarray}
 V\approx 
  -(6M_5^3)kw_\pi^2 c^2
  +|\lambda(\underline{\phi_1}^2-\mu^2)|^2+|m\underline{\phi_1}|^2 
\end{eqnarray}
which can be zero 
dependently on the parameters.

\section{Conclusion}\label{sc:conclusion}

We have studied supersymmetry breaking in 
a warped space model with constant boundary superpotentials,
hypermultiplet, boundary chiral supermultiplets,
compensator and radion multiplet.
We have presented the classical background solution
and have shown that all of the fields are determined unambiguously.

Dependence of the potential on $w_0$ and $w_\pi$
is significantly different from that of the potential in 
the model without the mixing between bulk and brane 
field equations \cite{Maru:2006id}.
The potential (\ref{nowo}) is independent of $w_0$.
In the situation we have considered where
the boundary chiral supermultiplets are only at $y=0$,
the constant superpotential at $y=\pi$ is required to be nonzero 
to stabilize
the modulus of the hypermultiplet.

For large negative $c$,
we have shown that the modulus of the hypermultiplet 
is stabilized at a finite value and that the radius is infinite.
It would be worth mentioning that 
large $|c|$ is closely related to 
flat space limit.
The bulk mass parameter $c$ should have large magnitude in
order to take a proper flat space limit $k\to 0$ as 
seen from the Lagrangian (\ref{lag}).
In Ref.\cite{Maru:2006id}, we found that
 there is
a similarity of hypermultiplet mass spectrum
between flat space case and $k\to 0$ limit of warped space case 
with fixed $ck$.
Infinite radius that we have obtained for large negative $c$
might be analogous to disappearance of the potential over $R$
in flat space case.

The radius stabilization has been studied also in the 
AdS$_4$ background where Scherk-Schwarz supersymmetry 
breaking is formulated. 
In models with nonzero superpotentials
\cite{Katz:2005wp}\cite{Katz:2006mv},
it has been found that hypermultiplets give positive 
contributions to the radion potential, contrary to the 
negative contributions from the gravity multiplet. 
This provides various patterns of radion potential. 
It would be interesting to study such a model with
mixing of equations of motion for bulk and brane fields.

\subsubsection*{Acknowledgments}
I thank Masud Chaichian, Nobuhito Maru and Norisuke Sakai
for useful discussions.
This work is supported by Bilateral exchange program between 
Japan Society for the Promotion of Science and the Academy of Finland.


\vspace*{10mm}


\end{document}